\documentclass[a4paper,11pt]{article}
\pdfoutput=1 

\usepackage{jinstpub} 
\usepackage{epsfig}
\usepackage{epstopdf}
\usepackage{graphicx}
\usepackage{subfigure}
\usepackage{hyperref}

\title{\boldmath Measurement of the ionization yield of nuclear recoils in liquid argon using a two-phase detector with electroluminescence gap}

\author[a,b]{A.~Bondar,}
\author[a,b]{A.~Buzulutskov,}
\author[b]{A.~Dolgov,}
\author[a]{E.~Grishnyaev,}
\author[a,b]{V.~Nosov,}
\author[a,b,1]{V.~Oleynikov,\note{Corresponding author.}}
\author[a,c]{S.~Polosatkin,}
\author[a,b]{L.~Shekhtman,}
\author[a,b]{E.~Shemyakina,}
\author[a,b]{A.~Sokolov}

\affiliation[a]{Budker Institute of Nuclear Physics, \\Lavrentiev ave. 11, Novosibirsk 630090, Russia}
\affiliation[b]{Novosibirsk State University,\\ Pirogova st. 2, Novosibirsk 630090, Russia}
\affiliation[c]{Novosibirsk State Technical University,\\ Karl Marx st. 20, Novosibirsk 630073, Russia}

\emailAdd{V.P.Oleynikov@inp.nsk.su}

\abstract{A measurement of ionization yields in noble-gas liquids is relevant to the energy calibration of nuclear recoil detectors for dark matter search and coherent neutrino-nucleus scattering experiments. In this work we further study the ionization yield of nuclear recoils in liquid Ar, using a two-phase detector with an electroluminescence gap and DD neutron generator. The ionization yields of nuclear recoils in liquid Ar were measured at 233 keV and electric fields of 0.56 and 0.62 kV/cm; their values amounted to 5.9 $\pm$ 0.8 and 7.4 $\pm$ 1 e$^-$/keV, respectively. The characteristic dependences of the ionization yield on energy and electric field were determined, while comparing the results obtained to those at lower energies and higher fields.}


\proceeding{Instrumentation for Colliding Beam Physics (INSTR17)\\
  27 February - 3 March, 2017\\
  Novosibirsk, Russia}

\begin{document}
\maketitle
\flushbottom

\section{Introduction}
The energy calibration of nuclear recoil detectors using liquid Ar and Xe detection media is of paramount importance in rare-event experiments
such as those of direct dark matter search and coherent neutrino-nucleus scattering \cite{NobleRev}. Such a calibration is usually performed by measuring the ionization yield and scintillation efficiency of nuclear recoils, using neutron elastic scattering off nuclei. While for liquid Xe there are ample experimental data on such yields \cite{LXeYield1,LXeYield2,LXeYield3}, little is known  about the ionization yield in liquid Ar.

The first results on the ionization yield of nuclear
recoils in liquid Ar were obtained just in the last three years: at lower energies, at 6.7 keV \cite{Joshi} and 17-57 keV \cite{Cao}, and at higher energies, at 80 and 233 keV \cite{IonYield14}. In the present work we continue the studies of the ionization yield in liquid Ar, using a new nuclear recoil detector as compared to our previous work \cite{IonYield14}, namely a two-phase Cryogenic Avalanche Detector (CRAD) with an electroluminescence (EL) gap. The present study complements the previous measurements and thus might be relevant to future dark matter search experiments \cite{ArDM,Darkside} and to thorough understanding of the ionization yield in liquid Ar.

The ionization yield measured in experiment is defined as follows:
\begin{equation}
\label{eq.1} Q_y=n_e/E_0 \,.
\end{equation}
Here  $n_e$ is the primary ionization charge, i.e. the number of ionization electrons escaping recombination with
positive ions; it depends on the energy deposited by a recoil
nucleus in the liquid ($E_0$) and on the electric field in the
liquid ($\mathcal{E}$). $n_e$  is always smaller than the initial
number of ion pairs produced in the liquid by a nuclear recoil
($N_i$). In the absence of a complete recombination model, it is
generally accepted that the following parametrization
works well \cite{NobleRev}:
\begin{equation}
\label{eq.2} n_e=\frac{N_i}{1 + k/\mathcal{E}} \,.
\end{equation}
where $k$ is a fitting parameter.

Equations \eqref{eq.1} and \eqref{eq.2} are valid for both electron recoils, induced by electron or gamma-ray irradiation, and nuclear recoils;
it is conventional to refer to the corresponding recoil
energy in units of keVee (electron-equivalent) and keVnr. The goal
of the present study is to measure $Q_y$ for nuclear recoils in
liquid Ar at 233 keV and different electric fields and to compare it to those of previous experiments.

The present study was performed in the course of the development of two-phase Cryogenic Avalanche Detectors (CRADs) of ultimate sensitivity for rare-event experiments \cite{CRADRev12,CryoMPPC15,CryoPMT15,CRADPropEL15,XRayYield16,CRADELGap17,ArXeN2Proc17,CPMTStudy17}.

\section{Experimental setup}

\begin{figure}[ht!]
\vfill \centering \subfigure[]{
    \includegraphics[width=0.40\linewidth]{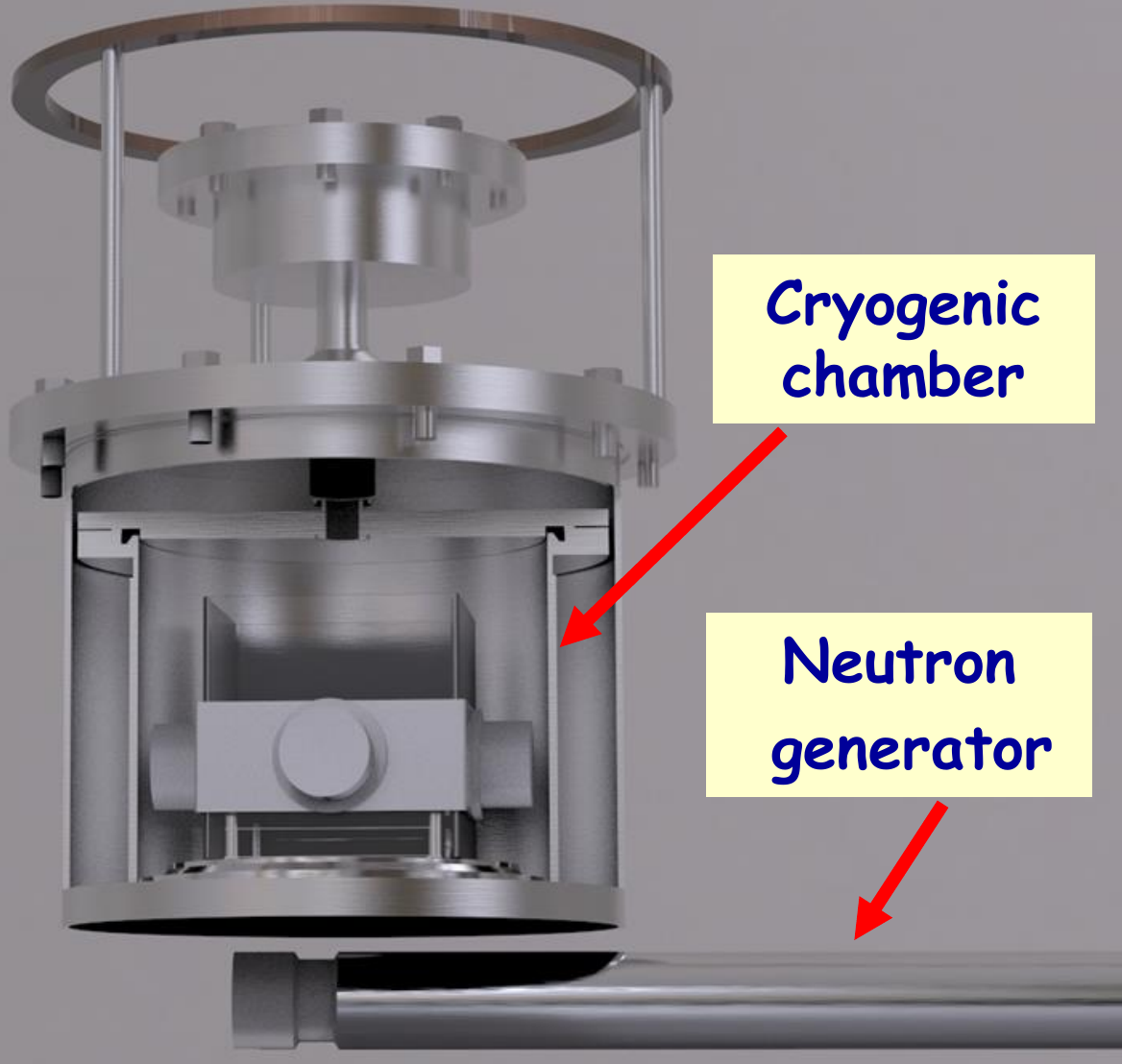} \label{image:setup_3d} }
\hfill
    \subfigure[]{
    \includegraphics[width=0.56\linewidth]{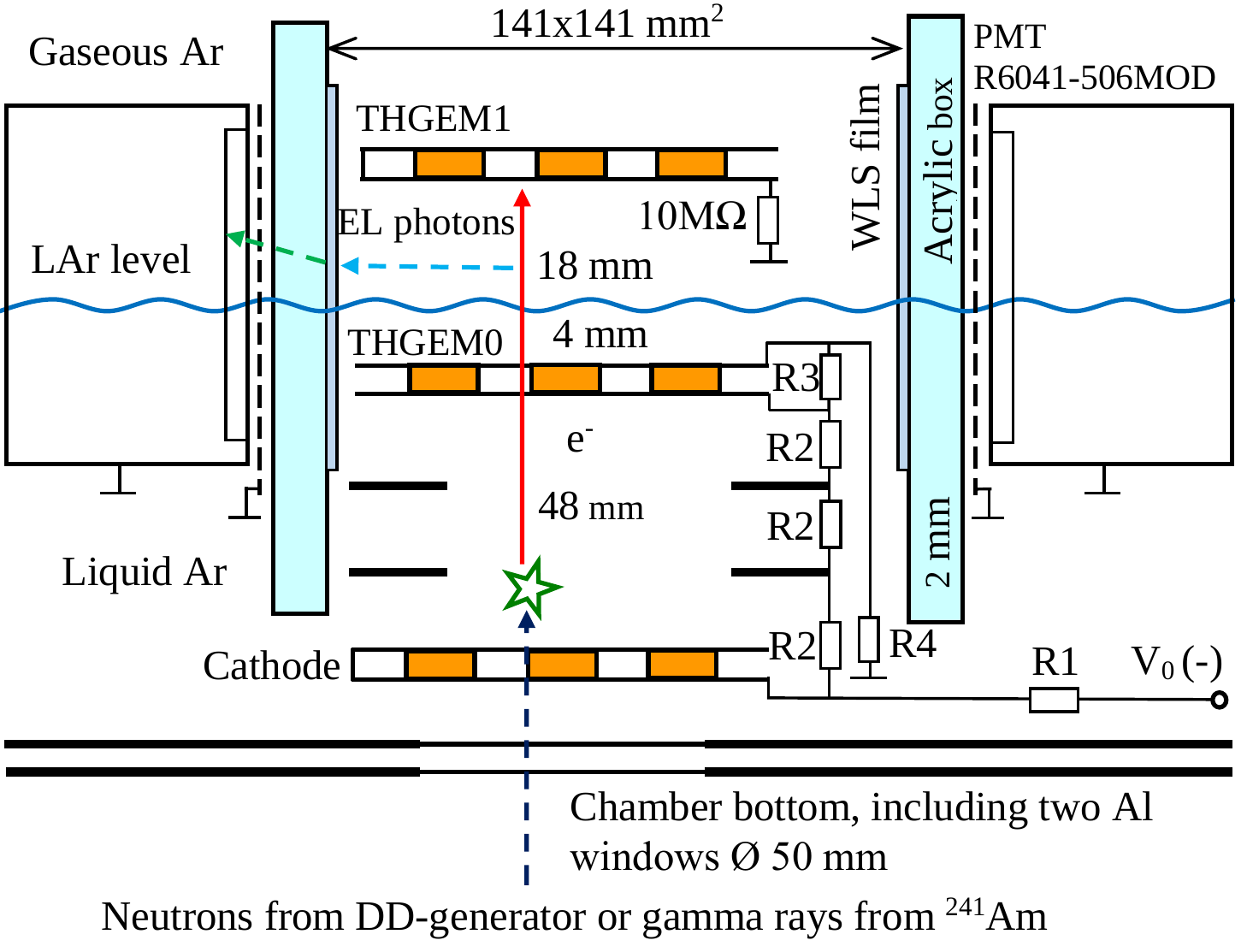} \label{image:setup_2d} }
\caption{\subref{image:setup_3d} 3d-view of the experimental setup; \subref{image:setup_2d} schematic view of the cryogenic chamber (not to scale).}
\end{figure}

In our previous work \cite{IonYield14}, a two-phase CRAD with charge (double-THGEM) readout was used to measure the ionization yield of nuclear recoils. In the present work,  we modified the nuclear recoil detector and the measurement conditions: see figures~\ref{image:setup_3d} and~\ref{image:setup_2d}. Firstly, we used a two-phase CRAD with optical readout, i.e with electroluminescence (EL) gap read out with cryogenic PMTs, which was supposed to have better energy resolution. Such a two-phase CRAD with EL gap was similar to that used in our recent studies of proportional electroluminescence in two-phase Ar \cite{CRADPropEL15,CRADELGap17}. Secondly, the active volume of the detector was increased by a factor of 6 compared to \cite{IonYield14}, by increasing the liquid Ar layer thickness, which allowed one to significantly reduce the measurement time. Thirdly, the measurements of the yield were performed at much lower electric fields in liquid Ar as compared to \cite{IonYield14}, at 0.56 and 0.62 kV/cm, which are more typical for dark matter search experiments \cite{ArDM,Darkside}.

The experimental setup included a vacuum-insulated cryostat with a nine-litre cryogenic chamber. The detector was operated in a two-phase mode in the equilibrium state, at a saturated vapour pressure of 1.0 atm and at a temperature of 87 K. The Ar was purified by an Oxisorb filter, providing an electron life-time in the liquid of $\ge$100 $\mu$s  \cite{CRADELGap17}.

The cryogenic chamber included a drift region (48 mm thick) and electron emission region (4 mm thick), in the liquid phase, and EL gap (18 mm thick), in the gas phase. All regions had an active area of
10$\times$10 cm$^2$.  The EL gap was viewed by four compact cryogenic 2-inch PMTs R6041-506MOD \cite{CryoPMT15}, located on the perimeter of the gap and separated from the high-field region by an acrylic box with WLS (wavelength shifter) films deposited in front of each PMT. The WLS films were needed to convert the VUV emission of pure Ar into visible light.

The primary ionization charge in the two-phase CRAD was produced by either X-rays from a $^{241}$Am source having a 59.5 keV line or by neutrons from a DD neutron generator. The primary ionization electrons produced in liquid Ar were first drifted towards and then emitted into the EL gap, where they were recorded via proportional electroluminescence using PMTs. The optical signals from the four PMTs were linearly summed and then amplified with a linear amplifier with a shaping time of 200 ns. The trigger was provided by the PMT signal itself at a certain detection threshold, well above the electronic and PMT noises. Since the electroluminescence signal was rather long, of several microseconds, the signal amplitude was defined as its pulse area.

To produce neutrons a specially designed  neutron generator was used
that continuously emitted monoenergetic neutrons (not collimated)
with the kinetic energy of 2.45 MeV obtained in the DD fusion
reaction \cite{NScatProp, NGen, NGen16}. The neutron flux was of the order
of 10$^4$ s$^{-1}$ over full solid angle.  The neutron generator
(operated at 80 kV and wrapped in a 1 cm thick Pb screen to suppress
bremsstrahlung gamma-rays) was placed underneath the two-phase CRAD
at a distance of about 10 cm from the active volume.

Other details of the experimental setup and measurement procedures were described elsewhere \cite{IonYield14,CRADPropEL15,CRADELGap17}.

\section{Experimental results}

\begin{figure}[ht]
	\center{\includegraphics[width=0.58\textwidth]{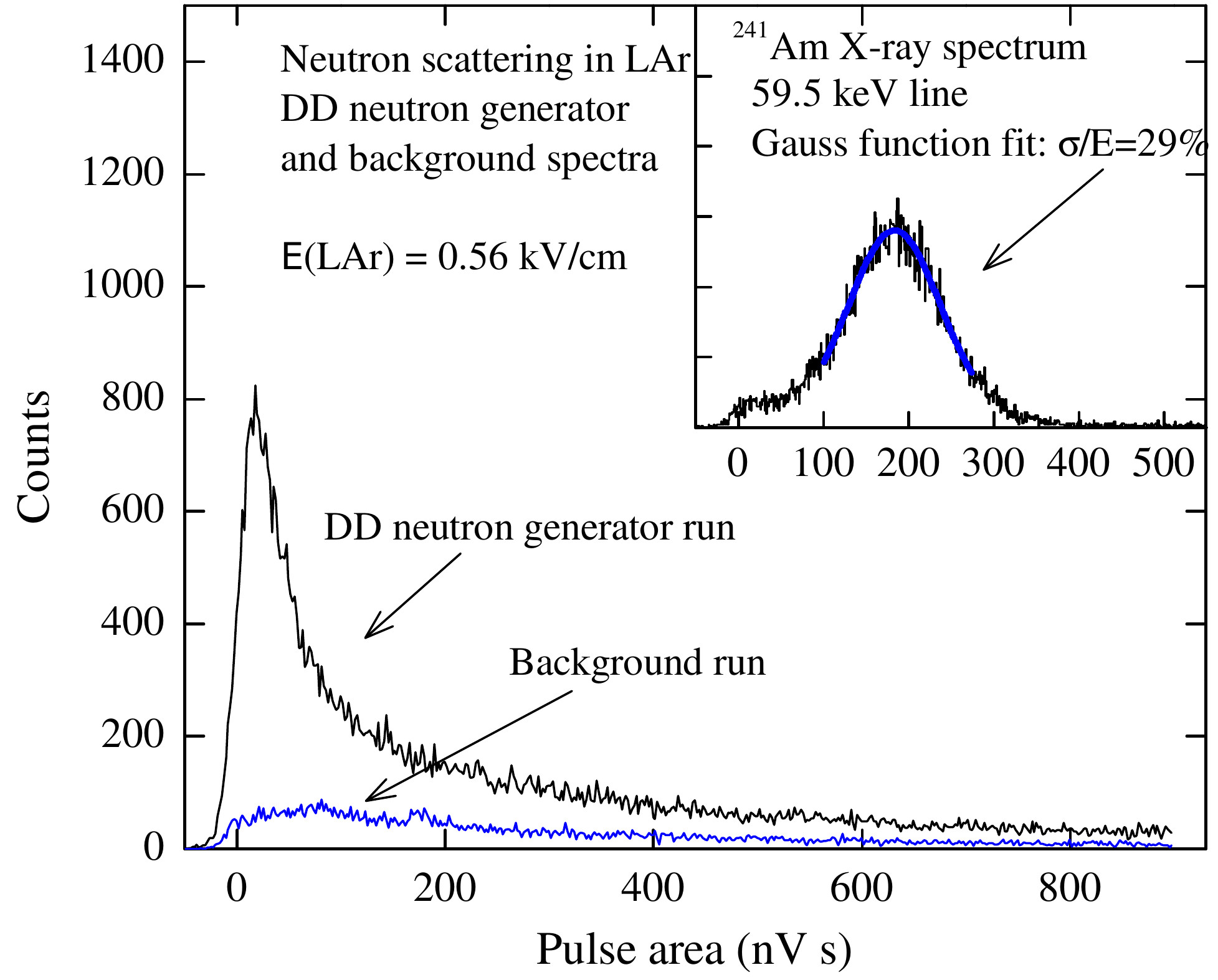}}
	\caption{Signal amplitude distribution of the two-phase CRAD in the measurement runs when the neutron generator was on (DD neutron generator run) and off (Background run). In the inset, the signal amplitude distribution induced by X-rays from a $^{241}$Am source in the calibration run is shown.}
	\label{SpectrRaw}
\end{figure}

The method to measure the ionization yield of nuclear recoils was similar to that of \cite{IonYield14}. The basic idea of the method is to compare the experimental amplitude spectrum of nuclear recoils expressed in terms of the primary ionization charge (e$^-$) to the theoretical spectrum expressed in terms of the nuclear recoil energy (keVnr). The experimental session was composed of the measurement runs during which the amplitude spectra were recorded: that with a neutron generator on and that with a neutron generator off (to measure the background contribution): see figure~\ref{SpectrRaw}. To obtain the real neutron scattering-induced spectrum, the latter should be subtracted from the former. To calibrate the amplitude scale in terms of the primary ionization charge, the detector was also irradiated with 59.5 keV X-rays from $^{241}$Am source in the calibration runs: see the inset in the figure. In addition in these calibration runs the amplitude resolution of the detector ($\sigma$/E) was measured: it amounted to 29\% and 23\% at 0.56 and 0.62 kV/cm, respectively.

In eq.~\eqref{eq.1} one should first determine $n_e$ from the spectra of figure~\ref{SpectrRaw}. For this the signal amplitude should be normalized to that of 59.5 keV peak and then
converted to the primary ionization charge. For the latter one
should know, in turn, the ionization yields for 59.5 keV X-rays in
liquid Ar, i.e. for electron recoils induced by X-ray absorption in liquid Ar; these were deduced from our previous work~\cite{XRayYield16}, for given electric fields in liquid Ar.

\begin{figure}[ht]
	\center{\includegraphics[width=0.58\textwidth]{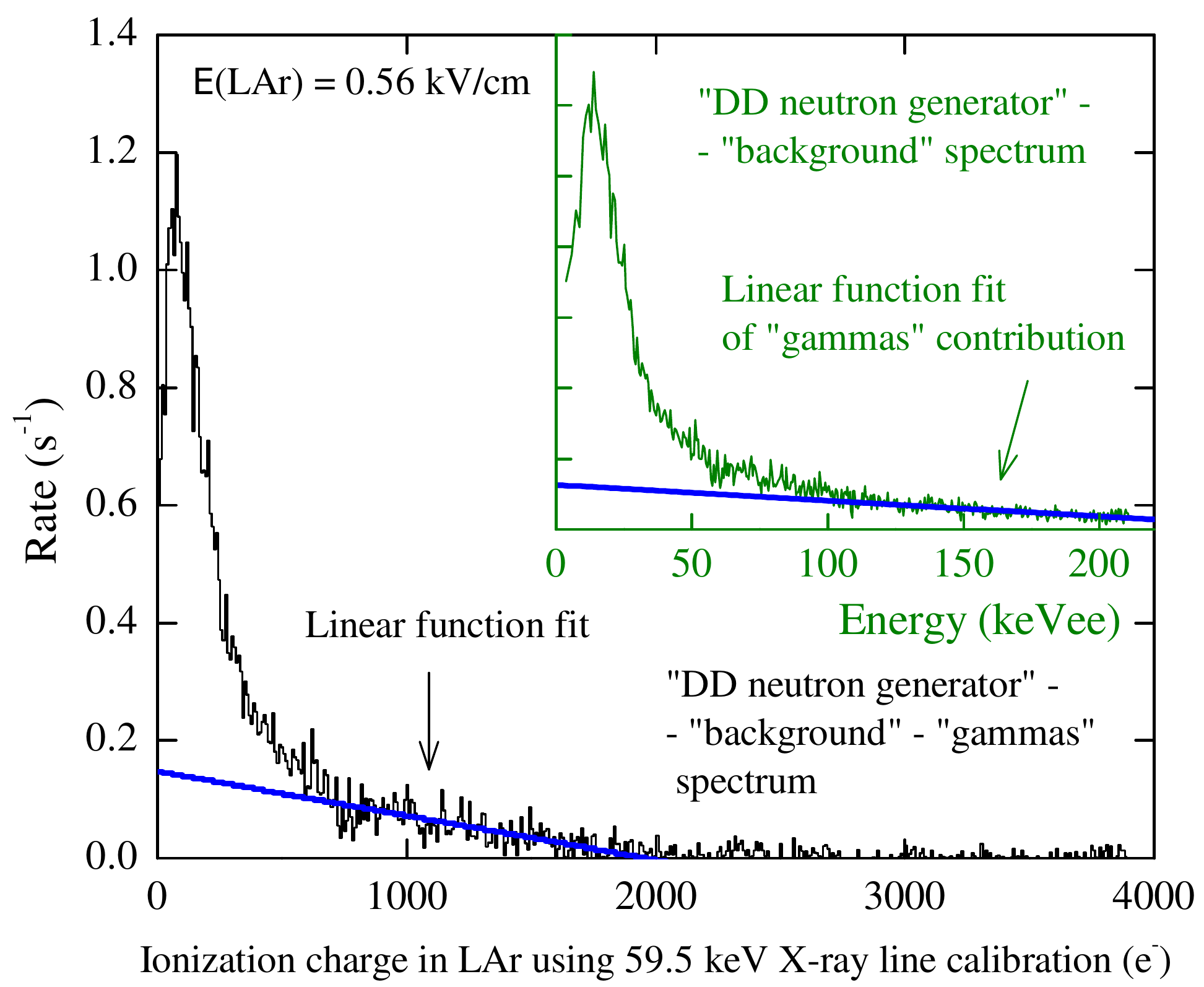}}
	\caption{Ionization charge distribution in liquid Ar induced by nuclear recoils due to neutron scattering, obtained from figure~\ref{SpectrRaw} using 59.5 keV X-ray line calibration and subtracting background-run and gamma-ray contributions. In the inset, the energy distribution is shown after subtracting the background-run contribution, but before subtracting the gamma-ray contribution, the latter being fitted by a linear decreasing function. Here the energy scale calibration was performed using the 59.5 keV X-ray line.}
	\label{SpectrNeutron}
\end{figure}

After subtracting the background-run contribution, the amplitude distribution still contains the gamma-ray background associated with ($n,\gamma$) reactions in surrounding materials. Similarly to  \cite{IonYield14}, this background was accounted for by fitting by a linear decreasing function: see the inset in figure~\ref{SpectrNeutron}.

Figure~\ref{SpectrNeutron} shows the desired ionization charge
spectrum in liquid Ar induced by nuclear recoils due to neutron
scattering; the spectrum is obtained from figure~\ref{SpectrRaw} using 59.5 keV X-ray line calibration and subtracting the background-run and gamma-ray contributions. The next step is to compare this experimental spectrum to the theoretical one.

\begin{figure}[ht!]
	\center{\includegraphics[width=0.58\textwidth]{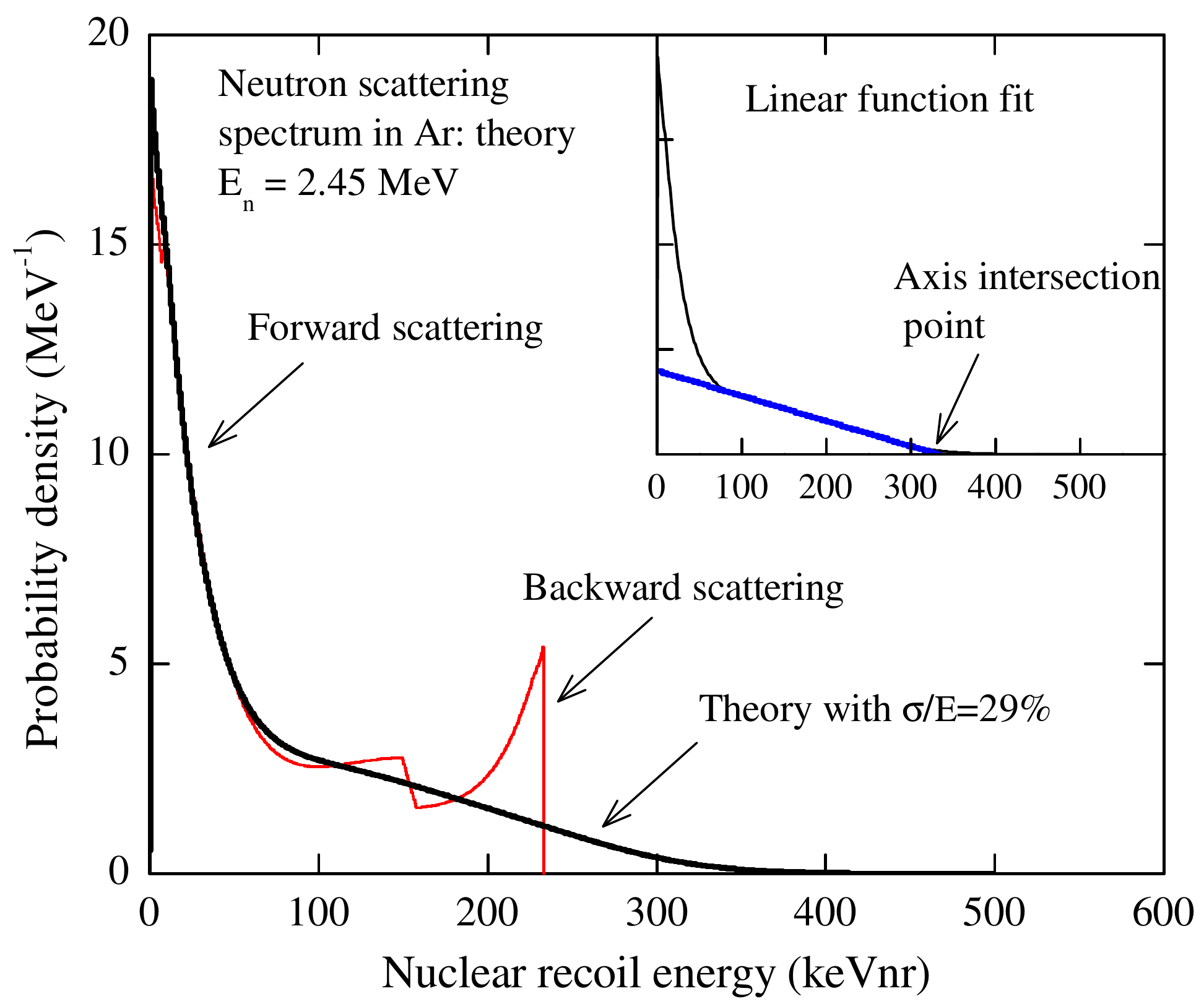}}
	\caption{Nuclear recoil energy distribution induced by scattering of
			neutrons with the energy of 2.45 MeV off $^{40}$Ar nuclei, computed
			theoretically (red line). The convolution of the theoretical
			spectrum with the energy resolution of the two-phase CRAD obtained
			in experiment (29\,\%), is also shown (black line). In the inset, the spectrum end-point is defined as an axis intersection point with a linear function fit of the backward scattering component.}
	\label{SpectrTheory}
\end{figure}

The theoretical  spectrum was computed similarly to  \cite{IonYield14}, using simulation code {\it
Scattronix} developed in \cite{NSpectr} and differential cross-sections of neutron scattering \cite{NData}: it is shown in
figure~\ref{SpectrTheory}.
The convolution of the theoretical spectrum with the energy resolution of the two-phase CRAD obtained in experiment (29\%) is also shown; just this spectrum should be
compared to the experimental one.  The backward scattering component (the shoulder) in both the experimental and theoretical spectra is well approximated by a linear function: see figure~\ref{SpectrNeutron} and the inset in figure~\ref{SpectrTheory}.
Similarly to  \cite{IonYield14}, the ionization yield at 233 keV was determined using the spectrum end-point (the horizontal axis intersection of the linear function fit): the ionization charge value (in the experimental spectrum) was divided by that of the nuclear recoil energy (in the theoretical spectrum).
The ionization yield values at two electric fields measured that way are as follows: $Q_y$~=~5.9~$\pm$~0.8~e$^-$/keV at 0.56 kV/cm and $Q_y$~=~7.4~$\pm$~1.0~e$^-$/keV at 0.62 kV/cm. These are presented in table~\ref{tab.1} along with the data of our previous work \cite{IonYield14}, showing also the appropriate statistical and systematic uncertainties.

\begin{table}
\caption{Ionization yields ($Q_y$) of nuclear recoils in liquid Ar measured at 233 keV.}
\label{tab.1}
\begin{center}
\begin{tabular}{|lccccc|}
\hline
Electric field & $Q_y$ & Statistical & Systematic & Total & Reference\\
(kV/cm)& (e$^-$/keV) & error & error & error & \\ [2mm]\hline
0.56  & 5.9 & 0.15 & 0.82 & 0.83 & this work\\
0.62  & 7.4 & 0.15 & 1.03 & 1.04 & this work\\
2.3  & 9.7 & 0.29 & 1.3 & 1.3 & \cite{IonYield14} \\
\hline
\end{tabular}
\end{center}
\end{table}

In addition to the ionization yield, the ionization efficiency (ionization quench factor) can be determined. The ionization efficiency is defined as the ratio of the ionization yield
of nuclear recoils ($Q_{y,nr}$) to that of electron  recoils ($Q_{y,ee}$), at the same energy:
$L_{ion}=Q_{y,nr}/Q_{y,ee}$. Combining the data of table~\ref{tab.1} and the data of
ionization yield of electron recoils from \cite{XRayYield16}, one can obtain that $L_{ion}$ amounts to 0.31~$\pm$~0.06 at 0.56 kV/cm and 0.37~$\pm$~0.07 at 0.62 kV/cm.

\section{Comparison with theoretical model and with other experiments}

Basically there are two theoretical models that can describe the
recombination effect in liquid noble gases and consequently the
energy and field dependence of the ionization yield of nuclear
recoils: that of Thomas-Imel \cite{TIModel1,TIModel2}, applicable mostly at lower energies, and that of Jaffe \cite{JaffeModel1,JaffeModel2}, applicable mostly at higher energies. For the ionization yield, the Thomas-Imel and Jaffe models predict the decreasing and increasing function of energy, respectively.

Accordingly, we used here the Jaffe model for data analysis in its compact form \cite{JaffeModel2}
\begin{equation}
\label{eq.6} n_e=\frac{N_i}{1+k_B(dE/dx)/\mathcal{E}} \,,
\end{equation}
resulting in the following equation for the ionization yield \cite{IonYield14}:
\begin{equation}
\label{eq.7}
Q_y=\frac{f}{[1+k_B(dE/dx)/\mathcal{E}][E_g+E_{ph}(N_{ex}/N_i)]} \,.
\end{equation}

\begin{figure}[ht!]
	\center{\includegraphics[width=0.6\textwidth]{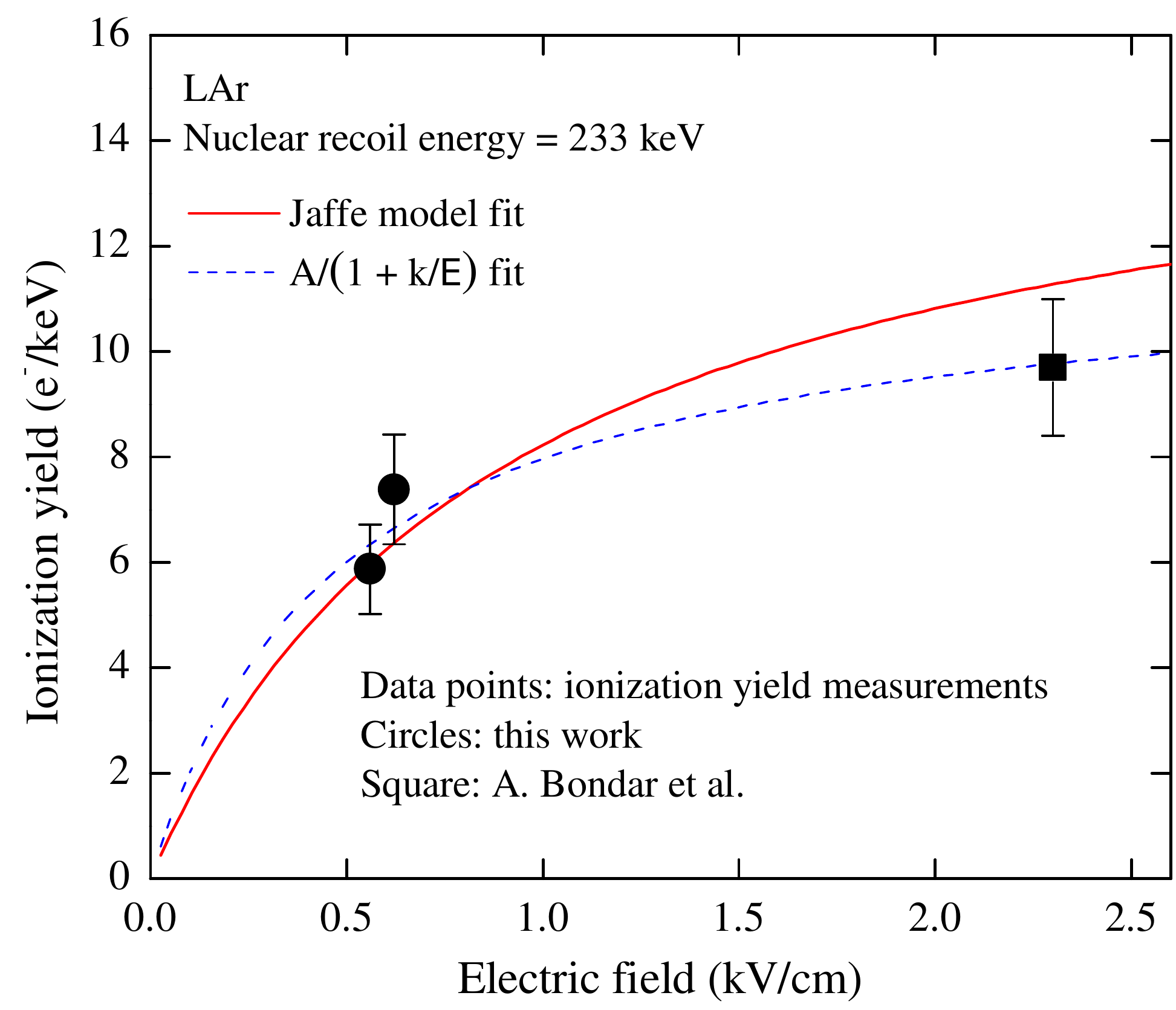}}
	\caption{Ionization yield of nuclear recoils in liquid Ar at 233
		keV as a function of the electric field, measured in this work and in A.~Bondar et al. \cite{IonYield14}. The theoretical model fits to the data points are also shown, in the frame of the Jaffe model and that of eq.~(\ref{eq.2}).}
	\label{IonYieldVsE}
\end{figure}

Here $k_B$ is a parameter; it is determined experimentally from the field dependence of the data (figure~\ref{IonYieldVsE}). Other parameters are taken the same as in \cite{IonYield14}. In particular, the stopping power for excitation and ionization, $dE/dx$, is calculated using a SRIM program \cite{SRIM}. $E_i=N_i E_g$ and $E_{ex}=N_{ex}E_{ph}$. Here $E_g\simeq$~14.2~eV is the band gap in liquid Ar; $E_{ph}$~=~9.7~eV is the average energy of scintillation photon. $N_{ex}/N_i$ is the ratio of the number of excitations to that of ionizations; its value is taken the same as in \cite{IonYield14}, $N_{ex}/N_i$~=~2, at which the energy dependence was better described within the Jaffe model. $f$ is the Lindhard factor; it is defined as the energy fraction transferred to ionization ($E_i$) and excitation ($E_{ex}$), $f=(E_i+E_{ex})/E_0$; it was calculated using the SRIM program \cite{SRIM}.

Figure~\ref{IonYieldVsE} shows the field dependence of the ionization yield at 233 keV, combining the data of the current and previous work \cite{IonYield14}. The data are described by the model of eq.~(\ref{eq.2}) or by Jaffe model of eq.~(\ref{eq.7}), where $k$ or $k_B$ is a free parameter. Both models fit well to the data points, resulting in $k$~=~0.48~$\pm$~0.26~kV/cm and  $k_B$~=~0.94~$\pm$~0.17~(V~mg)/(keV~cm$^3$). Note that the latter value is close to $k_B$~=~1.35~(V~mg)/(keV~cm$^3$) used in our previous work \cite{IonYield14}, where it was determined from the field dependence at 6.7 keV \cite{Joshi}.

\begin{figure}[ht!]
	\includegraphics[width=0.5\textwidth]{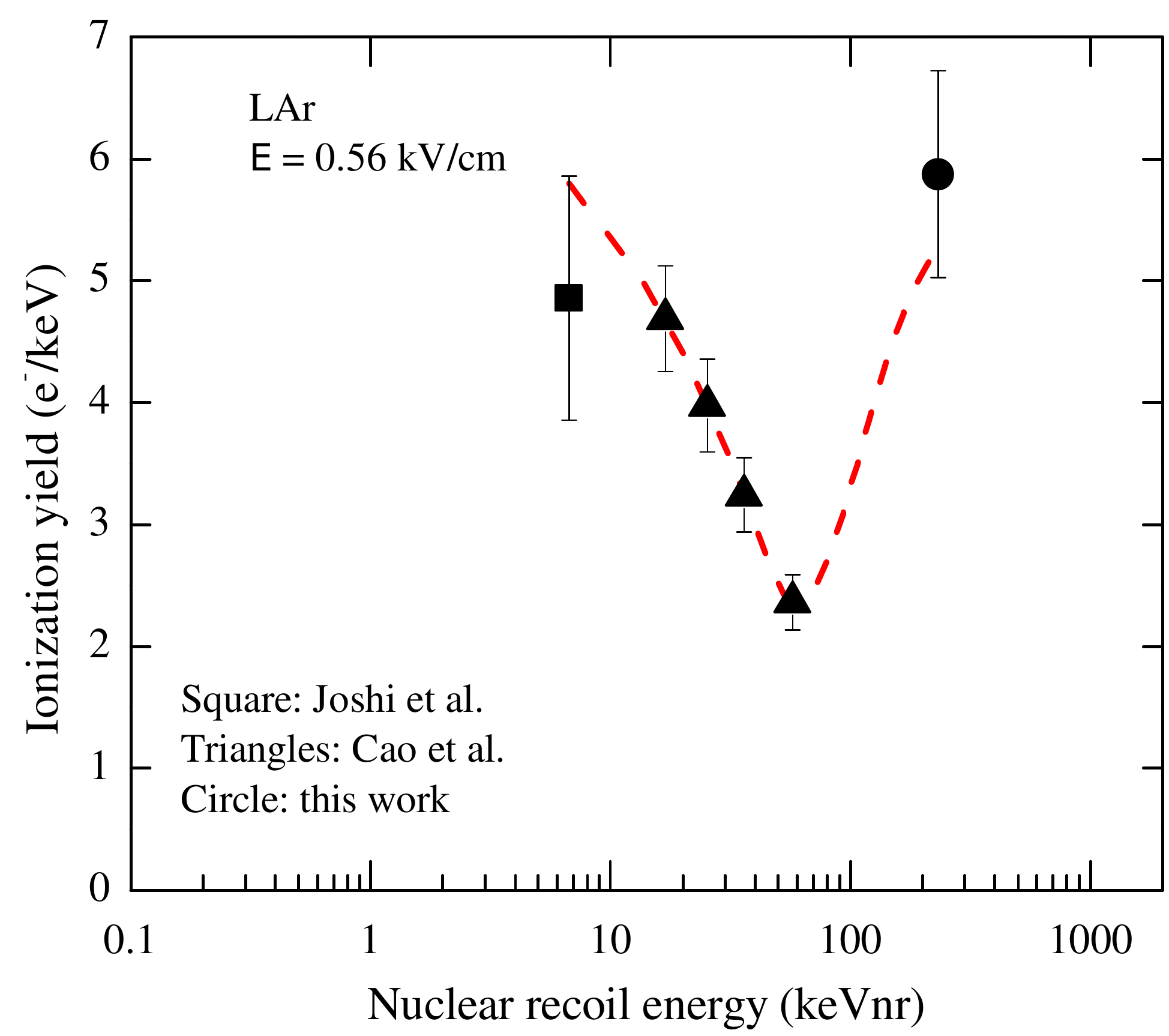}
	\hfill
	\includegraphics[width=0.5\textwidth]{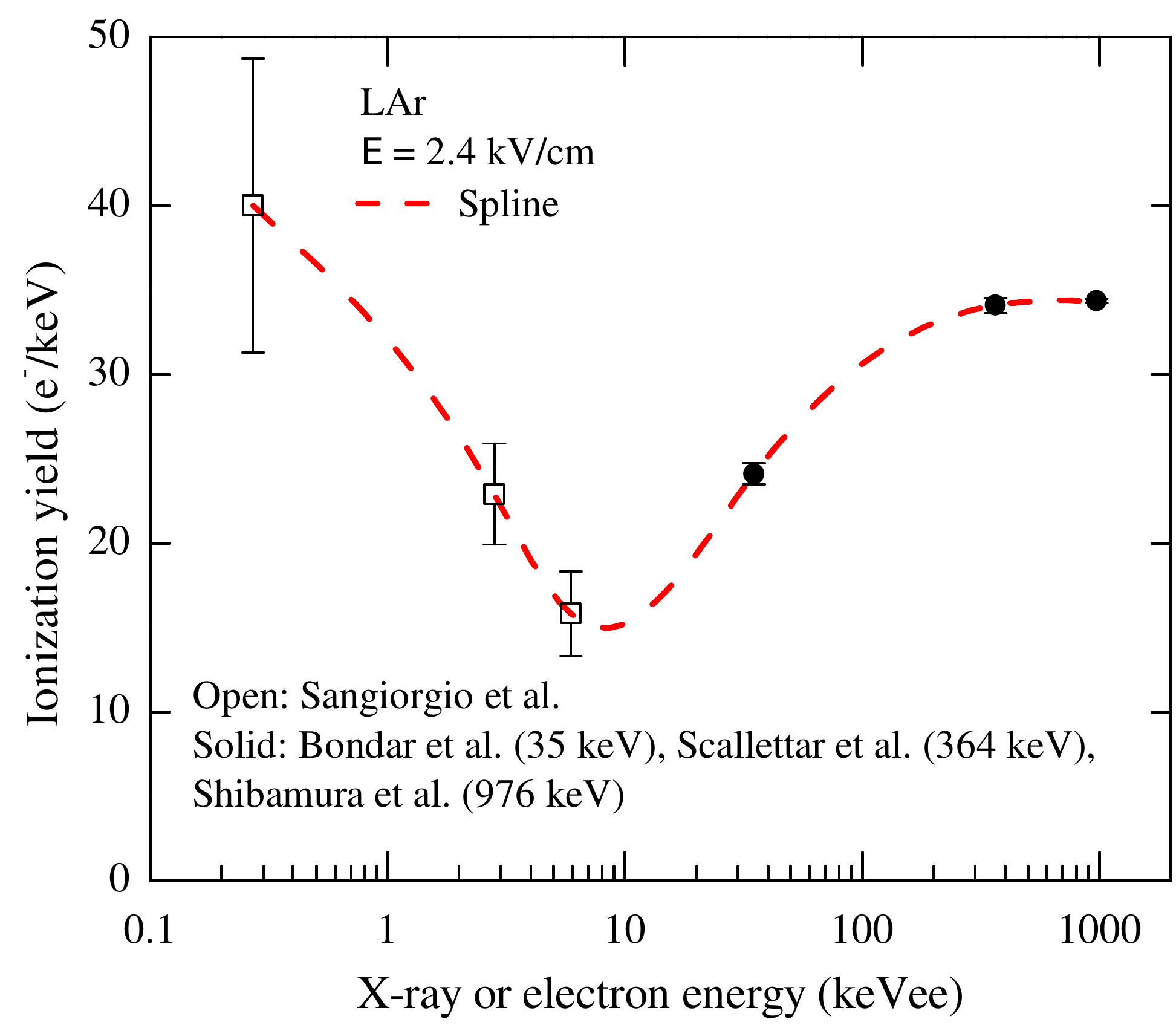}
	\\
	\parbox[t]{0.47\textwidth}{\caption{Ionization yield of nuclear recoils in liquid Ar at 0.56 kV/cm
 as a function of the energy, measured in Joshi et al. \cite{Joshi}, Cao et al. \cite{Cao} and this work.
The curve is drawn by eye.}
		\label{IonYield_combined_data}}
	\hfill
	\parbox[t]{0.47\textwidth}{\caption{Ionization yield of electron recoils in liquid Ar at 2.4 kV/cm
as a function of the energy, measured in Sangiorgio et al. \cite{Sangiorgio} (at 0.27 keV, 2.8 keV and 5.9 keV), in Bondar et al. \cite{XRayYield16} (at 35 keV), in Scallettar et al. \cite{Scalettar} (at 364 keV) and in Shibamura et al. \cite{Shibamura} (at 976 keV).
  The spline function fit for the ionization yield is also shown. The figure is taken from \cite{XRayYield16}.}
		\label{IonYield_electron}}
\end{figure}
Now we can try to determine the energy dependence of the ionization yield, combining the data of all three  experiments in the field: those at 6.7 keV \cite{Joshi}, 17-57 keV \cite{Cao} and 233 keV (this work). The data are extrapolated to the electric field value of 0.56 kV/cm; the result is shown in figure~\ref{IonYield_combined_data}.  One can see that the remarkable energy dependence is revealed: the ionization yield first decreases and then increases with energy, passing through a minimum. It is amazing that a similar energy dependence was observed for electron recoils in liquid Ar when combining the data from different experiments \cite{XRayYield16}: see figure~\ref{IonYield_electron}. On the other hand, the energy dependence for nuclear recoils in liquid Xe seems to be monotonic in the range of 1-300 keVnr \cite{Lenardo}.

\section{Conclusions}

In this work we further study the ionization yield of nuclear recoils in liquid Ar, using a two-phase detector with electroluminescence gap and DD neutron generator. The ionization yield in
liquid Ar at 233 keV was measured to be 5.9~$\pm$~0.8
and 7.4~$\pm$~1 e$^-$/keV at an electric field of 0.56 and 0.62 kV/cm, respectively; the ionization quench factor amounted to 0.31~$\pm$~0.06 and 0.37~$\pm$~0.07, respectively. The characteristic dependences of the ionization yield on the energy and electric field were determined, when comparing the results obtained to those of lower energies and higher fields. In particular, the remarkable energy dependence has been presumably revealed, when the ionization yield passes through a minimum with the energy increase.
The results of such a study are relevant to the energy calibration of liquid
noble gas detectors for dark matter search experiments and for thorough
understanding of the ionization yield in liquid Ar.

\acknowledgments
This study was supported by Russian Science Foundation
(project no. 14-50-00080); it was done within the
R\&D program for the DarkSide-20k experiment.


\end{document}